\documentclass[aps,prd,12point,twocolumn,nofootinbib,showpacs,superscriptaddress]{revtex4-2}
\def\theequation{\arabic{section}.\arabic{equation}}
\usepackage{psfrag}
\usepackage{subfigure}
\usepackage{color}
\usepackage{mathrsfs}
\usepackage{graphicx}
\usepackage{amssymb, bm}
\usepackage{amsmath, amsthm}
\usepackage{epstopdf}
\usepackage{hyperref}
\usepackage{enumerate}
\usepackage{longtable}
\usepackage{amsmath}
\usepackage[normalem]{ulem}
\usepackage[title]{appendix}

\newcommand{\be}{\begin{equation}}
\newcommand{\ee}{\end{equation}}

\definecolor{pinegreen}{rgb}{0.0, 0.47, 0.44}

\begin{document}
\def\theequation{\arabic{section}.\arabic{equation}}

\title{New phenomenology in the first--order thermodynamics of 
scalar--tensor gravity for  Bianchi universes}


\author{Julien Houle}
\email[]{ jhoule22@ubishops.ca}
\affiliation{Department of Physics \& Astronomy, Bishop's University, 
2600 College Street, Sherbrooke, Qu\'ebec, Canada J1M~1Z7
}
\author{Valerio Faraoni}
\email[]{vfaraoni@ubishops.ca}
\affiliation{Department of Physics \& Astronomy, Bishop's University, 
2600 College Street, Sherbrooke, Qu\'ebec, Canada J1M~1Z7
}

\begin{abstract} 

The phase space of Bianchi~I universes in vacuum Brans--Dicke gravity is 
analyzed in terms of physical variables. The behaviour of the solutions of 
the field equations near the fixed points (which are solutions of Einstein 
gravity) is compared with basic ideas of the recent first--order 
thermodynamics of scalar-tensor gravity, elucidating new phenomenology.

\end{abstract}



\maketitle

\section{Introduction}
\label{sec:1}
\setcounter{equation}{0}

Shortly after the introduction of general relativity (GR), researchers 
began looking for alternative theories of gravity, moved by pure curiosity 
about how things could be different in nature 
\cite{Weyl:1919fi,Eddington23}. 
More concrete 
motivation emerged with the birth of quantum field theory, when the 
question 
arose of how to reconcile the two biggest physics discoveries of the 
twentieth century, quantum mechanics and GR. The reason is that, as soon 
as one introduces the lowest--order quantum corrections to GR, one 
simultaneously causes deviations from it in the form of higher 
derivative terms in the field equations or extra degrees of freedom 
\cite{Stelle:1976gc,Stelle:1977ry}.  This situation does not change in 
string theory: the simplest string theory, the bosonic string, has a 
low--energy limit that reproduces not GR, but an $\omega=-1$ Brans--Dicke 
gravity \cite{Callan:1985ia,Fradkin:1985ys}. The prototype of alternative 
gravity is scalar--tensor gravity, which contains only a scalar degree of 
freedom $\phi$ in addition to the two massless spin two modes familiar 
from GR. The original Jordan--Brans--Dicke theory \cite{Brans:1961sx} was 
later generalized to wider scalar--tensor theories 
\cite{Bergmann:1968ve,Nordtvedt:1968qs, Wagoner:1970vr,Nordtvedt:1970uv} 
in which the ``Brans--Dicke coupling'' parameter became a function 
$\omega(\phi)$ of the scalar field $\phi$, which was also endowed with a 
potential $V(\phi)$.

There is independent motivation for the study of alternative theories of 
gravity coming from cosmological observations. The 1998 discovery, made 
with type Ia supernovae, that the present expansion of the universe is 
accelerated led to introducing overnight a completely {\em ad hoc} dark 
energy with very exotic properties (equation of state parameter close to 
$w\simeq-1$), of completely unkown nature and comprising approximately 
70\% of the energy content of the universe (see 
\cite{AmendolaTsujikawabook} for a review). A wide range of dark energy 
models have been proposed in the literature, but none is compelling and 
this state of affairs is deeply unsatisfactory from the theoretical point 
of view. For this reason, many cosmologists have turned to questioning 
whether, instead, we do not understand gravity on the largest 
(cosmological) scales and dark energy simply does not exist 
\cite{Capozziello:2003tk,Carroll:2003wy}. This idea had led to formulating 
and testing modified gravity models. Among a spectrum of possibilities, 
the most popular models belong to the so--called metric $f(R)$ gravity 
class (see \cite{Sotiriou:2008rp,DeFelice:2010aj,Nojiri:2010wj} for 
reviews). Metric $f(R)$ gravity contains only one extra massive, 
propagating, scalar degree of freedom and, therefore, falls into the wider 
category of scalar--tensor gravity 
\cite{Sotiriou:2008rp,DeFelice:2010aj,Nojiri:2010wj}. Even Starobinski 
inflation \cite{Starobinsky:1980te}, the first scenario of inflation 
and the one currently favored by observations \cite{Planck2}, is based on 
quadratic corrections to the Einstein--Hilbert action and is 
ultimately a scalar--tensor theory.

In the past decade, the problem of finding the most general scalar--tensor 
theory expressed by field equations that are at most of second order led 
to the rediscovery and intense study of the older Horndeski gravity 
\cite{Horndeski}. This sought--for property was found to belong not to 
Horndeski gravity but to the more general Degenerate Higher Order 
Scalar--Tensor (DHOST) theories, a subclass of higher order gravities in 
which a degeneracy condition brings the order of the field equations back 
to two ({\em e.g.}, \cite{H1, H2, H3, GLPV1,GLPV2, DHOST1, DHOST2,DHOST3, 
DHOST4, DHOST5, DHOST6, DHOST7,Creminellietal18}, see 
\cite{Langloisetal18,DHOSTreview1,DHOSTreview2} for reviews).

Given the wide spectrum of scalar--tensor gravities (not to mention other 
alternatives richer in propagating degrees of freedom that are difficult 
to identify and count \cite{deRham:2023brw}), what is the role of GR in 
this landscape? A proposal is well--known in the context of emergent 
gravity, in which the field equations can be deduced as an emergent or 
collective property of underlying degrees of freedom and are not 
fundamental. The seminal paper by Jacobson \cite{Jacobson:1995ab} derived 
the Einstein equations of GR from purely thermodynamical considerations, 
an idea referred to as ``thermodynamics of spacetime''. This feat was 
repeated with quadratic $f(R)$ gravity producing a new picture: GR 
is somehow a state of ``thermal equilibrium'' of gravity, while 
alternative theories correspond to entropy generation and to excited 
thermal states \cite{Eling:2006aw}. This view has been very influential 
and has generated a large literature but, unfortunately, no substantial 
progress has been made since its early days. Specifically, 
the ``temperature of gravity'' (or other order parameter) has not been 
identified and no equation describing the relaxation of alternative 
gravity to GR has been proposed, although there are reasons to believe 
that such phenomena could have occurred in the early universe 
\cite{Damour:1993id,Damour:1992kf}.

Recently a new proposal has been advanced, known as the first--order 
thermodynamics of scalar--tensor gravity 
\cite{Faraoni:2018qdr,Faraoni:2021lfc,Faraoni:2021jri, 
Giusti:2022tgq,Faraoni:2023hwu}: it has nothing to do with Jacobson's 
thermodynamics of spacetime, although some of its ideas share the same  
spirit. It is a far--reaching analogy (but, still, only an analogy) 
between the description of the effective stress--energy tensor of the 
scalar degree of freedom $\phi$ of scalar--tensor gravity and the 
stress--energy tensor of a 
dissipative fluid. By writing the field equations of scalar--tensor 
gravity as effective Einstein equations, the contributions of $\phi$ and 
of its first and second derivatives form an effective stress--energy 
tensor 
$T_{ab}^{(\phi)}$ which has the form of a dissipative fluid stress--energy 
tensor \cite{Faraoni:2018qdr} (this fact has been known for a long time 
for special theories or special geometries \cite{Madsen:1988ph,Pimentel89} 
and has been recently recognized also for ``viable'' Horndeski gravity 
\cite{Quiros:2019gai,Giusti:2021sku}).  Specifically, if the scalar field 
gradient $\nabla^a \phi$ is timelike and future--oriented 
\cite{Giusti:2022tgq}, its normalized version
\be 
u^a  \equiv \frac{ \nabla^a \phi}{\sqrt{ -\nabla^c\phi \nabla_c \phi}} 
\label{4velocity}
\ee 
can 
be seen as the four--velocity of an effective fluid with  
stress--energy tensor of the form 
\be 
T_{ab}^{(\phi)} = \rho u_a u_b + P 
h_{ab} +\pi_{ab} + q_a u_b + q_b u_a \,,\label{decomposition} 
\ee 
where 
$\rho$ is an effective energy density, $P$ is an effective isotropic 
pressure, $\pi_{ab}$ is an effective anisotropic, trace--free, stress 
tensor, and $q^a$ is an effective heat flux density. Here $h_{ab} \equiv 
g_{ab} +u_a u_b$ is the Riemannian three-space metric seen by observers  
comoving with this fluid (with ${h^a}_b$ the projector onto this 3-space), 
while $\pi_{ab}$ and $q^a$ are purely spatial: 
\be 
h_{ab}u^a = h_{ab} u^b 
= 0 \,, \quad \pi_{ab}u^a =\pi_{ab}u^b=0 \,, \quad q_c u^c=0 \,. 
\ee 
The fact that this $T_{ab}^{(\phi)}$ has the dissipative fluid form 
contains no physics: the decomposition~(\ref{decomposition}) holds true 
for any symmetric two--index tensor \cite{Faraoni:2023hwu}. However, when 
one takes seriously this dissipative fluid structure and tries to apply to 
it Eckart's \cite{Eckart40} theory of dissipative fluids 
\cite{Faraoni:2018qdr,Faraoni:2021lfc,Faraoni:2021jri} one discovers that, 
miraculously, Eckart's constitutive relation
\be 
q_a = -{\cal K} h_{ab} \left( \nabla^b {\cal T}+ T \dot{u}^b \right) 
\label{Eckart} 
\ee 
holds.  Here ${\cal T}$ is the temperature of the dissipative fluid, 
${\cal K}$ is 
the thermal conductivity, and $\dot{u}^a$ is the fluid four--acceleration. 
Equation~(\ref{Eckart}) is nothing but the relativistic generalization of 
Fourier's law with the addition of an inertial term proportional to the 
four--acceleration, which takes into account the fact that heat is a form 
of energy and its transport contributes to the energy flux in a way 
that was absent in pre--relativistic physics \cite{Eckart40}. The 
unexpected fact that this relation holds for the effective $\phi$--fluid 
makes it possible to define the product ${\cal KT}$ for scalar--tensor 
gravity.

Let us refer, for simplicity, to ``first--generation'' ({\em i.e.}, 
pre--Horndeski) scalar--tensor gravity: in the Jordan frame, the 
gravitational sector of the theory is described by the action\footnote{ We 
adopt the notation of Ref.~\cite{Waldbook}: the signature of the 
spacetime metric $g_{ab}$ is ${-}{+}{+}{+}$ and units are used in which 
the the speeed of light $c$ and 
Newton's constant $G$ are unity. The Ricci tensor and Ricci scalar are 
\be
R_{ab} \equiv  R^c_{\,\,acb} = \partial _c\Gamma^c_{ba} - \partial 
_b\Gamma^c_{ca} + 
\Gamma^c_{cd}\Gamma^d_{ba} - \Gamma^c_{bd}\Gamma^d_{ca} \,,
\ee
\be
R \equiv {R^a}_a = \partial _c\Gamma^c_{ba} - \partial _b\Gamma^c_{ca} + 
\Gamma^c_{cd}\Gamma^d_{ba} - \Gamma^c_{bd}\Gamma^d_{ca} \,.
\ee
}  
\be
S_\mathrm{ST} = \frac{1}{16\pi}\int d^4x\sqrt{-g}\left[ \phi R- 
\frac{\omega(\phi)}{\phi}\nabla^c\phi\nabla_c\phi-V(\phi) \right] 
\label{action}
\ee 
where $\phi > 0$ is the Brans-Dicke scalar (approximately equivalent to 
the inverse of the effective gravitational coupling strength $G^{-1}$), 
the function $\omega(\phi)$ (which was a strictly constant parameter in 
the original Brans-Dicke theory) is the “Brans-Dicke coupling”, and 
$V(\phi)$ is a scalar field potential (absent in the original Brans-Dicke 
theory). When $\nabla^a \phi$ is timelike and future--oriented, the 
product of effective thermal conductivity and effective temperature is 
found to be \cite{Faraoni:2021lfc,Faraoni:2021jri}
\be
{\cal KT}= \frac{ \sqrt{ -\nabla^c \phi \nabla_c \phi} }{8\pi \phi} \,. 
\label{KT}
\ee
It is apparent that GR, reproduced for $\phi=$~const.~$>0$, corresponds to 
${\cal KT}=0$. It is rather intuitive that, when extra degrees of freedom 
with respect to GR are excited, gravity is in some sense excited and 
``hotter'' than the GR state in which the extra degrees of freedom are 
absent. 
The ``temperature of gravity'' ${\cal T}$ is, in this sense, a temperature 
{\em relative to the GR state of equilibrium}. 

The first--order thermodynamics of scalar--tensor gravity has been 
extended \cite{Giusti:2021sku} to ``viable'' Horndeski gravity, {\em 
i.e.}, to the subclass in which gravitational waves propagate at the speed 
of light, and applied to various situations in cosmology and other 
contexts \cite{Giardino:2022sdv, Miranda:2022wkz,Faraoni:2022gry, 
Miranda:2022uyk, Faraoni:2022doe, Faraoni:2022fxo, Faraoni:2022jyd, 
Faraoni:2023hwu, Giardino:2023qlu} (see \cite{ Giardino:2023sw} for a 
recent review). Two basic qualitative ideas emerge from these studies: 
near spacetime singularities gravity is ``hot'', {\em i.e.}, it diverges 
from GR; the expansion of the three--space perceived by comoving 
observers (with 3--metric $h_{ab}$) ``cools'' gravity, bringing it closer 
to GR. These ideas have been tested against various situations of physical 
interest or mathematical convenience (for which it is possible to draw 
analytically definite conclusions). Here we continue this program. While 
it was natural to apply the first--order thermodynamics of scalar--tensor 
gravity to Friedmann--Lema\^itre--Robertson--Walker (FLRW) cosmology 
\cite{Giardino:2022sdv,Miranda:2024dhw}, here we extend the description to 
anisotropic Bianchi universes. For simplicity, we confine ourselves to 
Brans--Dicke gravity and to the simplest anisotropic cosmologies described 
by spatially flat Bianchi~I geometries. To describe the dynamics we resort 
to a phase space view and, contrary to the literature that we are aware 
of, we choose as phase space variables the average Hubble function $H$, 
the scalar field $\phi$, and its time derivative $ \dot{\phi}$ that have a 
direct physical meaning instead of other variables obtained by various 
non--linear combinations of physical ones (for this reason we cannot avail 
ourselves of existing phase space analyses).

\section{Field equations and Bianchi~I geometry}
\label{sec:2}
\setcounter{equation}{0}

The (Jordan frame) vacuum field equations obtained by varying the 
action~(\ref{action}) with respect to the inverse metric $g^{ab}$ and the 
scalar $\phi$ are
\begin{eqnarray}
R_{ab} - \frac{1}{2} \, g_{ab}R & = &  
\frac{\omega}{\phi^2}(\nabla_a\phi\nabla_b\phi - 
\frac{1}{2}g_{ab}\nabla_c\phi\nabla^c\phi)  \nonumber\\
&&\nonumber\\
&\, &  +\frac{1}{\phi}\left( \nabla_a\nabla_b\phi - g_{ab}\square \phi 
\right) - 
\frac{V}{2\phi}g_{ab} \,,\nonumber\\
&& \label{fe1}
\end{eqnarray}
\be
\square\phi = \frac{1}{2\omega + 3} \left(  \phi \, \frac{dV}{d\phi} - 2V - 
\frac{d\omega}{d\phi} \, \nabla^c\phi\nabla_c\phi \right) \,.\label{fe2}
\ee
In the following we assume $ V(\phi) \geq 0 $, constant Brans--Dicke 
coupling $\omega$, and  $ 2\omega+3 >0$ to avoid phantom scalar fields 
$\phi$. 

The line element of a Bianchi~I universe is  
\be
ds^2 = -dt^2 + A^2(t)dx^2 + B^2(t)dy^2 + C^2(t)dz^2 \label{BianchiI}
\ee
in Cartesian comoving coordinates $\left( t,x,y,z \right)$, where $A(t), 
B(t)$, and $C(t)$ are the scale factors associated with the three orthogonal  
spatial directions. This anisotropic universe has 
average scale factor $a(t) \equiv \left( ABC \right)^{1/3}$ and average Hubble parameter
\be
H \equiv \frac{ \dot{a} }{a} = \frac{1}{3} \left( \frac{ \dot{A} }{A}  + 
\frac{\dot{B} }{B}  + \frac{ \dot{C}}{C} \right)  
\equiv \frac{1}{3} \left( H_A + H_B + H_C \right) \,,
\ee 
where an overdot denotes differentiation with respect to the cosmic time $t$, 
$H_i \equiv \dot{A}_i/A$ (where $A_i= A,B$, or $C$), and 
\begin{eqnarray}
H^2 &=& \frac{1}{9} \left( H_A^2 + H_B^2 + H_C^2 + 2H_A H_B + 2H_A H_C 
\right.
\nonumber\\
&&\nonumber\\
&\, & \left. + 2H_B H_C \right) \,.
\end{eqnarray}
The shear scalar $\Sigma$ for this scalar--tensor gravity, computed 
in terms of $\phi$ and its derivatives, is given by 
\cite{Faraoni:2018qdr} 
\begin{eqnarray}
\sigma & \equiv &  \sqrt{\frac{1}{2} \, \sigma_{ab}\sigma^{ab}} 
\nonumber\\
&&\nonumber\\
&=& \left( -\nabla^e\phi\nabla_e\phi \right)^{-3/2} 
\left\{\frac{1}{3}(\nabla_a\nabla_b\phi \nabla^a\phi\nabla^b\phi)^2  
\right. \nonumber\\
&&\nonumber\\
&\, & + \frac{1}{2}(\nabla^e\phi\nabla_e \phi)^2 \left[ \nabla^a\nabla^b\phi 
\nabla_a\nabla_b\phi- 
\frac{1}{3}(\square\phi)^2 \right]  
- \left( \nabla^e\phi\nabla_e\phi \right)
\nonumber\\
&&\nonumber\\
&\, & \left. \times \left(\nabla_a\nabla_b\phi\nabla^b\nabla_c\phi 
- \frac{1}{3}\square\phi\nabla_a\nabla_c\phi\right) 
\nabla^a\phi\nabla^c\phi\right\}^{1/2} 
\,,\nonumber\\
&&
\end{eqnarray}
where $\sigma_{ab}$ is the shear tensor \cite{Waldbook}. 
For clarity, we use the shear variable $\Sigma \equiv \frac{1}{2} \sigma^2$  which, in the Bianchi~I 
geometry~(\ref{BianchiI}), 
assumes the 
form 
\begin{eqnarray}
\Sigma & = & \frac{1}{6A^2B^2C^2} \left( B^2C^2\Dot{A}^2 + 
A^2C^2\Dot{B}^2 + A^2B^2\Dot{C}^2 \right. \nonumber\\
&&\nonumber\\
&\, & \left. - ABC^2\Dot{A}\Dot{B} - AB^2C\Dot{A}\dot{C} - 
A^2BC\Dot{B}\Dot{C}\right) \nonumber\\
&&\nonumber\\
&=& \frac{B^2C^2\Dot{A}^2 + A^2C^2\Dot{B}^2 + A^2B^2\Dot{C}^2}{6A^2B^2C^2} 
\nonumber\\
&&\nonumber\\
&\, &  - 
\frac{C\Dot{A}\Dot{B}  + B\Dot{A}\dot{C} + A\Dot{B}\Dot{C}}{6ABC} \nonumber\\
&&\nonumber\\
&=& \frac{1}{6} \left( H_A^2 + H_B^2 + H_C^2 - H_AH_B - H_AH_C - H_BH_C \right) \nonumber\\
&&\nonumber\\
&=&  \frac{1}{4} \left(H_A^2 + H_B^2 + H_C^2\right) - \frac{3}{4}H^2 
\nonumber\\
&&\nonumber\\
&=& \frac{3H^2}{2} - \frac{1}{2}\left( H_A H_B + H_A H_C + H_B H_C \right) 
\,.
\end{eqnarray}
$\Sigma$ vanishes if and only if all the components of $\sigma_{ab}$ 
vanish \cite{Ellis71}, in which case the Bianchi geometry reduces to a 
FLRW one.

The only non--vanishing Christoffel symbols of the 
geometry~(\ref{BianchiI}) are
\begin{eqnarray}
\Gamma_{xx}^t &=& A \dot{A} \,, \quad \Gamma_{yy}^{t} = B \dot{B} \, ,\quad \Gamma_{zz}^t =  
C \dot{C} \,, 
\nonumber\\
\Gamma_{tx}^{x} &=& \Gamma_{xt}^{x}=  \frac{\dot{A}}{A} \,, \quad 
\Gamma_{ty}^{y}  =\Gamma_{yt}^{y} =  \frac{\dot{B}}{B} \,, \nonumber\\
\Gamma_{tz}^{z} &=& \Gamma_{zt}^{z}=  \frac{\dot{C}}{C} \,, 
\end{eqnarray}
while the non--vanishing components of the Ricci tensor are
\begin{eqnarray}
R_{tt} &=&  - \left( \frac{ \ddot{A}}{A} + \frac{ \ddot{B}}{B}  + 
\frac{ \ddot{C} }{C} \right) \,,\\
&&\nonumber\\
R_{xx} &=&  \frac{ A B C \ddot{A} + \left( A C \dot{B} + AB \dot{C} \right) 
\dot{A} }{B C} 
\,,\\
&&\nonumber\\
R_{yy} &=& \frac{A B C \ddot{B} + \left( BC \dot{A}  + A B\dot{C} \right) \dot{B} }{AC} \,, 
\\
&&\nonumber\\
R_{zz} &=&  \frac{ABC \ddot{C} + \left( BC \dot{A} + AC \dot{B} \right) \dot{C}}{AB}  \,,
\end{eqnarray}
and the Ricci scalar reads
\begin{eqnarray}
R &=&  2 \left( \frac{\Ddot{A}}{A} + \frac{\Ddot{B}}{B} + 
\frac{\Ddot{C}}{C} + H_A H_B + H_B 
H_C + H_AH_C \right) \nonumber\\
&&\\
&=&  6 \left( \Dot{H} + 2H^2  + \frac{2\Sigma}{3} \right) \,.
\end{eqnarray}

The time--time component of the Brans--Dicke field equations~(\ref{fe1}) 
is
\be
H^2 = \frac{\omega}{6} \left( \frac{\Dot{\phi}}{\phi} \right)^2 + 
\frac{V}{6\phi} + \frac{2\Sigma}{3} - H\,  \frac{\Dot{\phi}}{\phi} 
\,,\label{eq:1}
\ee
while the spatial components read
\begin{gather}
-\omega ABC\Dot{\phi}^2 + ABCV\phi - 2\phi^2 \left( A\Ddot{B}C + 
A\Dot{B}\Dot{C} + 
AB\Ddot{C} \right) \nonumber\\
-2ABC\phi\Ddot{\phi} -  2\phi\Dot{\phi} \left( A\Dot{B}C+AB\Dot{C} \right) 
= 0 \,,
\end{gather}

\begin{gather}
-\omega ABC\Dot{\phi}^2 + ABCV\phi - 2\phi^2 \left( \Ddot{A}BC + 
\Dot{A}B\Dot{C} + 
AB\Ddot{C} \right) \nonumber\\
-2ABC\phi\Ddot{\phi} - 2\phi \Dot{\phi} \left( \Dot{A}BC+AB\Dot{C} \right) 
= 0 \,,
\end{gather}

\begin{gather}
-\omega ABC\Dot{\phi}^2 +  ABCV\phi - 2\phi^2 \left( \Ddot{A}BC + 
\Dot{A}\Dot{B}C + A\Ddot{B}C \right) \nonumber\\
-2ABC\phi\Ddot{\phi} - 2\phi \Dot{\phi} \left( \Dot{A}BC+A\Dot{B}C \right) 
= 0 \,.
\end{gather}
and the trace of the field equation~(\ref{fe1}) is
\be
\Dot{H} = -\frac{\omega}{6} \left( \frac{\Dot{\phi}}{\phi} \right)^2 + 
\frac{V}{3\phi} -  \frac{2\Sigma}{3} - 2H^2 - \frac{ \left( \Ddot{\phi} + 
3H\Dot{\phi} \right)}{2\phi}  \,.\label{trace}
\ee

Equation~(\ref{fe2}) for the scalar field is  
\be
\Ddot{\phi} + 3H\Dot{\phi} + \frac{ \phi V' - 2V   }{2\omega + 3} =0 \,, 
\label{eq:3}
\ee
where a prime denotes differentiation  with respect to $\phi$. Combining 
these three equations yields  
\be
\Dot{H} = -\frac{\omega}{2} \left( \frac{\Dot{\phi}}{\phi} \right)^2 - 
2\Sigma + 
2H \, \frac{\Dot{\phi}}{\phi} 
+ \frac{ \left( \phi V' - 2V  \right)}{2\phi \left( 2\omega + 3 
\right)}  \,. \label{eq:combination}
\ee

By inserting Eq.~(\ref{eq:3}) for the scalar field $\phi$ 
into~(\ref{trace}) and combining the result with Eq.~(\ref{eq:1}), one 
obtains
\begin{gather}
\Dot{H} = - 3H^2 -\frac{H \Dot{\phi}}{\phi} +\frac{ \phi V'(\phi)+(2\omega 
+ 1)V(\phi)}{2\phi(2\omega + 3)} \,. \label{secondequation}
\end{gather}

To summarize, the field equations to be solved for the scalar field 
$\phi(t)$ and the Hubble function $H(t)$ are Eq.~(\ref{eq:3}) and 
Eq.~(\ref{secondequation}), respectively. Once these quantities are known, 
Eq.~(\ref{eq:1}) gives the shear $\Sigma$.

\section{Phase space}
\label{sec:3}
\setcounter{equation}{0}

Let us discuss the phase space of Bianchi~I cosmologies in vacuum 
Brans--Dicke gravity, where the dynamics is due entirely to the 
Brans--Dicke scalar $\phi$.  We use the variables $ \left( H, \phi, 
\Dot{\phi} \right)$ which are physical: the Hubble parameter $H$ is a 
cosmological observable (although its actual value is subject to a very 
significant tension \cite{Verde:2019ivm, DiValentino:2021izs}), while 
$\phi$ is the extra scalar degree of freedom of scalar--tensor gravity in 
addition to the two spin zero massless modes of GR contained in the metric 
$g_{ab}$. The strength of the gravitational coupling $G\simeq \phi^{-1}$ 
is measured directly by Cavendish experiments and its time variation ({\em 
i.e.}, $\dot{G}$ and, consequently,  $\dot{\phi}$) is subject to 
observational constraints \cite{Willbook,Will:2014kxa}.  By contrast, much 
of the existing 
literature on Bianchi 
cosmologies uses variables which are complicated functions of $H, \phi$, 
and $\dot{\phi}$ and do not have direct physical interpretation. Although 
they may make the study of the phase space dynamics more convenient from 
the formal 
point of view, they have no direct physical meaning. Here we want to 
interpret the dynamics and 
compare it with the first--order thermodynamics of spacetime, therefore we 
must use physical variables.

Equation~(\ref{eq:1}) yields the shear 
\begin{equation}\label{eq:5}
    \Sigma \left( H, \phi, \dot{\phi} \right) = \frac{3}{2} \left[ H^2 + 
H \, \frac{\dot{\phi}}{\phi}  - \frac{\omega}{6} \left( 
\frac{\dot{\phi}}{\phi} \right)^2 - 
\frac{V}{6\phi} \right]  
\end{equation}
as a function of the  three phase space variables, therefore $\Sigma$ is 
not an independent variable, although it will be relevant in our analysis. 


With our choice of variables the fixed points in the phase space, if they 
exist, have necessarily the form $\left( H, \phi, \dot{\phi} \right)  = 
\left( H_0 ,\phi_0 , 0 \right)$, with $H_0$ and $\phi_0 >0$ constants. 
They 
are solutions of the Einstein equations located on the ``GR plane'' 
$\dot{\phi}=0$ of the phase space identified by constant scalar field, the 
condition that reproduces GR.

Using the notation $ V (\phi_0) \equiv V_0$ and $ V'(\phi_0)\equiv V'_0$, 
Eq.~(\ref{eq:5}) that must be satisfied by the fixed points 
gives
\begin{equation}\label{eq:6}
\Sigma_0 = \frac{3}{2} \left( H_0^2 - \frac{V_0}{6\phi_0} \right) \,,
\end{equation}
while the mix of Eqs.~(\ref{trace}) and~(\ref{eq:3}) gives, using $\Dot{H} 
= \omega' = 0$, 
\begin{equation}\label{eq:7}
 H_0^2 =  \frac{V_0}{6\phi_0} - \frac{\Sigma_0}{3}  + \frac{\phi_0 V'_0 - 
2V_0 }{4\phi_0 \left( 2\omega + 3 \right)}
\end{equation}
and the scalar field equation degenerates into 
\begin{equation}\label{eq:8}
V'_0 = \frac{2V_0}{\phi_0}
\end{equation}
(which is non--negative since $V\geq 0$ and $\phi>0$) reducing 
Eq.~(\ref{eq:7}) to
\begin{equation}\label{eq:9}
H_0^2 =  \frac{V_0}{6\phi_0} - \frac{\Sigma_0}{3} \,.
\end{equation}
Comparing Eqs.~(\ref{eq:6}) and~(\ref{eq:9}), one obtains
$ \Sigma_0 =  0$: as expected, the shear vanishes at the fixed points, 
which have $A=B=C\equiv a(t)$ and  
the average Hubble function $H(t)$ coincides with the FLRW one. 
Equation~(\ref{eq:7}) 
then becomes $H_0^2 =  V_0/ \left( 6\phi_0 \right)$, or 
 \begin{equation}\label{eq:10}
H_0 =  \pm\sqrt{\frac{V_0}{6\phi_0}} =\pm \sqrt{ \frac{ V_0'}{12}} \,.
\end{equation}
The degenerate fixed points corresponding to $V_0 = V'_0 = H_0 = 0$ are Minkowski 
spaces, while those corresponding to $V_0 > 0$, $V'_0 > 0$, and 
$H_0 =  
\pm\sqrt{\frac{V_0}{6\phi_0}}$ are de Sitter spaces.\footnote{With an 
abuse of nomenclature, 
we refer to the fixed points with $H_0>0$ as ``expanding de Sitter 
spaces'' and to those with 
$H_0<0$ as ``contracting de Sitter spaces''.} When $\phi$ becomes  
a constant $\phi_0$ and $V(\phi_0) \equiv V_0$ is positive, the theory of 
gravity reduces to GR with a positive cosmological constant $\Lambda=V_0$.
 
In the study of exact solutions of the field equations, one sometimes find 
solutions $\left( H(t), \phi(t), \dot{\phi}(t) \right)$ with $\phi(t) \to 
0^{+}$ at late times $t\to +\infty$: these are pathological as the 
effective gravitational coupling $G_\mathrm{eff} \to +\infty$. The line 
$\phi=0$ in the ``GR plane'' $\dot{\phi}=0$ corresponds to singularities 
at which $G_\mathrm{eff}$ changes sign, but it does so by going through 
$G_\mathrm{eff}=\infty$ (a similar situation occurs with conformally 
coupled scalar fields \cite{Starobinsky2,Hrycyna:2012zc}). Exact solutions 
with these 
properties are unphysical and cannot be regarded as GR solutions, even 
though the first--order thermodynamics of scalar--tensor gravity does not, 
strictly speaking, indicate a pathology or a gross deviation from GR in 
this situation (see Appendix~\ref{Appendix:A}).

\subsection{Stability of the equilibrium points}
\label{subsec:3A}

Let us examine the stability of the fixed points with respect to 
homogeneous perturbations described by 
\begin{eqnarray}
\phi(t) &=& \phi_0 + \delta\phi(t) \,,\\
&&\nonumber\\
H(t) &=&  H_0 + \delta H(t) \,.
\end{eqnarray}
Evolution equations for the perturbations $\delta \phi(t) , \delta H(t)$ 
are obtained from Eq.~(\ref{eq:3}) for the scalar 
field $\phi$ and Eq.~(\ref{secondequation}) for $H$. 

Let us begin with the stability of the scalar field. By expanding the 
scalar field potential, $V(\phi) \simeq  V_0 + V'_0 \, \delta\phi$, and 
using 
the zero--order field equation~(\ref{eq:3}), one obtains the 
linearized equation for $\delta \phi$
\begin{gather}
    \delta \Ddot{\phi} +  3 H_0 \delta\Dot{\phi} + 
\omega_0^2 \, \delta\phi = 
0 \,,  \label{HO}
\end{gather}
where 
\be
\omega_0^2 \equiv \frac{\phi_0V_0'' - V'_0}{2\omega + 3} 
= \frac{\phi_0V_0'' - 2V_0/\phi_0}{2\omega + 3} = 
\frac{\phi_0V_0'' - 12H_0^2}{2\omega + 3} \,. \label{omeganot}
\ee
The {\em ansatz} 
\be
\delta\phi(t) = 
\delta_0 \, \mbox{e}^{\alpha t} 
\ee
with $\delta_0$ and $\alpha$ constants  yields the algebraic 
equation
\begin{gather}
    \alpha^2 + 3H_0 \alpha + \omega_0^2 = 0 \label{algebraic}
\end{gather}
with roots
\begin{gather}
\alpha_{(\pm)} =\frac{ -3H_0 \pm \sqrt{9H_0^2 - 4\omega_0^2}}{2} 
\equiv \frac{1}{2} \left( -3H_0 \pm \sqrt{\Delta} \right)
\,\label{alpha}
\end{gather}
and two modes $\delta\phi_{(\pm)}(t) = \delta_0 \, 
\mbox{e}^{\alpha_{(\pm)}t} $ (this applies if $\alpha_{(\pm)} \neq 
0$; the case $\alpha_{(\pm)}=0$ is discussed separately).

\begin{itemize} 

\item If $\Delta<0$, corresponding to $\omega_0^2> 9H_0^2/4$ and 
$\omega_0$ real, 
then 
\be
\delta\phi_{(\pm)}(t) =  \delta_0  \, \mbox{e}^{-\frac{3H_0 t}{2}} \, 
\mbox{e}^{\pm \frac{i}{2} \,  \sqrt{|\Delta|} \, t  } \,:
\ee
the second exponential in the right--hand side oscillates while, as 
$t\to+\infty$, the first exponential 
diverges if $H_0<0$ and decays if $H_0>0$ (it remains constant if 
$H_0=0$). In this case the fixed point is stable if $H_0 \geq 0$ and 
unstable if $H_0<0$.

\item If $\Delta=0$, corresponding to $\omega_0^2= 9H_0^2/4$ (and 
$\omega_0$ 
real), the scalar field perturbation is simply $\delta \phi (t) = \delta_0 
\, \mbox{e}^{-\frac{3H_0 t}{2} } $ and is stable if $H_0 \geq 0$, unstable 
if $H_0<0$.

\item If $\Delta>0$, corresponding to $\omega_0^2< 9H_0^2/4$ then it is 
convenient to write
\be
\alpha_{(\pm)} = \frac{3H_0}{2} \left( -1 \pm 
\sqrt{ 1- \left( \frac{2\omega_0}{3H_0} \right)^2} \, \right) \,.
\ee
If $\omega_0$ is real, corresponding to $ \phi_0 V_0'' \geq 12H_0^2$, then 
$\sqrt{ 1- \left( \frac{2\omega_0}{3H_0} \right)^2} <1$ and 
$ -1 \pm \sqrt{ 1- \left( \frac{2\omega_0}{3H_0} \right)^2}<0$, hence 
$\alpha_{(\pm)}<0$ if $H_0>0$, or $ \alpha_{(\pm)}=0$ if $H_0=0$ 
(corresponding to a Minkowski fixed point), or $\alpha_{(\pm)}>0$ if 
$H_0<0$. Fixed points with $H_0 \geq 0$ are stable while those with 
$H_0<0$ 
are unstable.

If instead $\omega_0$ is imaginary, $\omega_0 = i|\omega_0 |$, 
corresponding to $\phi_0V_0'' <12H_0^2$, then we have 
$\sqrt{ -1 + \left( \frac{2\omega_0}{3H_0} \right)^2} > 0 $ and  
$-1-\sqrt{ 1- \left( \frac{2\omega_0}{3H_0} \right)^2} <0$. The mode 
$\delta \phi_{(+)}$ is unstable ({\em i.e.}, $\alpha_{(+)}>0$) if 
$H_0>0$, while the other mode 
$\delta\phi_{(-)} $ is unstable ({\em i.e.}, $\alpha_{(-)}>0$) if 
$H_0<0$. In short, when $\omega_0$ is imaginary and $H_0\neq 0$ there is 
always a unstable mode and the fixed point is unstable.

If instead $\omega_0$ is imaginary and $H_0=0$ (Minkowski fixed point), 
the equation for the scalar field perturbations reduces to
\be
\delta\ddot{\phi} -|\omega_0^2| \delta \phi =0 \,,
\ee
which describes an unstable inverted harmonic oscillator. 

\end{itemize}

An exception not included in the previous discussion is the situation 
in which $V= 0$ and $H_0=0$, $\omega_0=0$ corresponding to a  
Minkowski space. In this case, Eq.~(\ref{HO}) reduces to 
$\ddot{\phi}=0$, which has a linear solution and this Minkowski space 
is unstable.

Let us consider now the perturbation $\delta H(t)$. 
Using the zero--order equations, Eq.~(\ref{secondequation}) gives the 
linearized equation of motion for $\delta H$ 
\begin{gather}
   \delta \Dot{H} + 6H_0 \delta H =   
-\frac{H_0}{\phi_0} \, \delta \Dot{\phi} 
+ \frac{[V''_0\phi_0 + \left( 2\omega + 1 
\right) \frac{V_0}{\phi_0}]}{ 2\phi_0 \left( 2\omega 
+ 3 \right) } \, \delta\phi
\end{gather} 
and Eq.~(\ref{omeganot}) yields 
\begin{gather}
 \delta   \Dot{H} + 6H_0\delta H=   
-\frac{H_0}{\phi_0} \,\delta \Dot{\phi} 
+\frac{ \left( \omega_0^2 + 6H_0^2 \right)}{2\phi_0}\,  \delta\phi \,.
\end{gather}
Using the explicit form  $\delta\phi_{(\pm)}(t)$ of the 
scalar field perturbation gives 
\begin{gather}
\delta \Dot{H} + 6H_0\delta H = \beta_{(\pm)} \delta\phi \,,
\end{gather}
where
\be
\beta_{(\pm)} = \frac{ \omega_0^2 + 6H_0^2 
-2\alpha_{(\pm)} H_0 }{2\phi_0} \,.
\ee
The solution of this inhomogeneous ordinary differential equation is 
\begin{eqnarray}
\delta H_{(\pm)} (t) &=& \frac{\beta_{(\pm)} }{ \alpha_{(\pm)} + 6H_0} 
\, \delta \phi + C \, \mbox{e}^{ -6H_0 t }  \nonumber\\
&&\nonumber\\
&=& \frac{ \left( H_0 -\alpha_{(\pm)} \right) }{2\phi_0 } \, \delta\phi   
+ C \, \mbox{e}^{  -6H_0 t }  \,,
\end{eqnarray}
where $C$ is an integration constant. The perturbation $ \delta H(t)$ 
diverges for $H_0<0$ regardless of the behaviour of the scalar field 
perturbation $\delta\phi$.

To summarize:
\begin{itemize}

\item Contracting de Sitter spaces are always unstable fixed 
points, which can be understood as the effect of anti--friction in the 
(anti--)damped harmonic oscillator equation~(\ref{HO}). 

\item Expanding de Sitter fixed points are stable if $ \phi_0 V_0'' > 
12H_0^2$ and unstable if $ \phi_0 V_0'' <12H_0^2$. 

\item Minkowski fixed points are (marginally) stable if $\omega_0$ is 
real (corresponding to $ \phi_0 V_0'' \geq 12H_0^2$) and unstable 
otherwise. The exception is the Minkowski space obtained for $V\equiv 
0, H_0=0$, which is unstable.

\end{itemize}

Let us consider now the ratio of the shear variable to the expansion 
variable $\Sigma/H_0^2$, which quantifies the amount of anisotropy and the 
departure from GR (remember that the fixed points, when they exist, all 
lie in the GR plane $\dot{\phi}=0$ and $\Sigma=0$).  Since the equilibrium 
points are isotropic de Sitter spaces, the shear~(\ref{eq:5}) is purely 
perturbative and given by 
\begin{eqnarray}
\Sigma &=& \delta \Sigma = \frac{3}{2} \left[ \left( H_0 + \delta H 
\right)^2 +  \frac{ \delta \Dot{\phi} }{\phi_0 + \delta\phi} \left( H_0 + 
\delta 
H \right) \right.\nonumber\\
&&\nonumber\\ 
&\, &   \left. - \frac{\omega}{6} \left( 
\frac{ \delta \Dot{\phi} }{\phi_0 + \delta\phi} \right)^2 
- \frac{ \left( V_0 + V'_0 \delta\phi \right) }{ 6\left( \phi_0 + 
\delta\phi 
\right)} \right] 
\end{eqnarray}
which, to first order, reduces to 
\begin{gather}
    \frac{\delta\Sigma}{H_0^2} = \frac{3}{2H_0}  \left[2\delta H + 
 \frac{\delta\Dot{\phi}}{\phi_0}  - \frac{H_0}{\phi_0} \, 
\delta\phi\right] \,.
\end{gather} 
Inserting the solution for the perturbations $\delta\phi, \delta H$ 
into this equation and using Eq.~(\ref{algebraic}) to express $\omega_0^2$  
yields 
\begin{gather}
    \frac{\delta\Sigma}{H_0^2} = \frac{3C}{H_0} \, \mbox{e}^{-6H_0 t} \,,
\end{gather}
(Since $\Sigma \simeq \delta \Sigma \geq 0$, one deduces that $ 
\mbox{sign}\left( C\right) = \mbox{sign}\left( H_0 \right)$.) 

The ratio $\Sigma/H_0^2$ vanishes as $t \to +\infty$ for all solutions 
that are perturbations of expanding de Sitter fixed points, and diverges 
in the same late--time limit near contracting de Sitter fixed points.

Let us now examine the significance of these results with respect to the 
first--order thermodynamics of spacetime of 
Refs.~\cite{Faraoni:2018qdr,Faraoni:2021lfc,Faraoni:2021jri, 
Giusti:2022tgq,Faraoni:2023hwu}. We emphasize that the following 
discussion is meaningful only because physical phase space variables 
$\left( H, \phi, \dot{\phi} \right) $ have been chosen at the outset.

de Sitter solutions with non--constant scalar field, which are possible in 
scalar--tensor gravity but not in GR, have been discussed in 
Ref.~\cite{Faraoni:2022doe}.

\subsection{Comparison with the first-order thermodynamics of 
scalar--tensor gravity}

In Brans--Dicke theory, the temperature of gravity relative to the GR 
state of equilibrium \cite{Faraoni:2018qdr,Faraoni:2021lfc,Faraoni:2021jri, 
Giusti:2022tgq,Faraoni:2023hwu} is given by Eq.~(\ref{KT}). For linear 
homogeneous perturbations of the fixed points, it reads  
\begin{gather}
    \mathcal{KT} = \frac{|\alpha_{(\pm)}|  |\delta\phi_{(\pm)}|}{ 8\pi 
\phi_0}  \geq 0 \,.
\end{gather}
The fixed points, which lie in the GR plane with $\dot{\phi}=0$ clearly
correspond to $\mathcal{KT} = 0$. ${\cal KT}$ assumes  the same form as in 
a FLRW  universe, but the solution $\phi(t)$ is, in general, 
different in Bianchi~I and in FLRW universes.

If the orbit of a solution in the $\left( H, \phi, \dot{\phi} \right)$ 
phase space lies near an expanding de Sitter fixed point and is attracted 
to it, the anisotropic three-space expands,  and the 
solution converges to the zero--temperature state of equilibrium, while 
$\Sigma/H_0^2 \to 0$ and this three-space isotropizes. The cooling of 
gravity (${\cal KT}\to 0$) is indeed another way of saying that GR is  
a late--time attractor of the dynamics. The situation is not so trivial, 
however, because there are exceptions for $\phi_0 V_0'' < 12H_0^2$. In 
this case three--space still expands exponentially but the de Sitter 
fixed 
point nearby is a repellor. It is still the case that $H_0>0$ and 
$\Sigma/H_0^2\to 0$ as $t \to +\infty$. How do we understand this 
situation in the light of scalar--tensor thermodynamics? The answer comes 
from examining the equation ruling the approach to/departure from the GR 
equilibrium state derived in 
Refs.~\cite{Faraoni:2018qdr,Faraoni:2021lfc,Faraoni:2021jri, 
Giusti:2022tgq,Faraoni:2023hwu}
\be
\frac{d\left( {\cal KT}\right)}{d\tau} = 8\pi \left( {\cal KT} \right)^2 
-\Theta {\cal KT} +\frac{ \Box\phi}{8\pi \phi} \,,\label{KTevolution}
\ee
where $\tau$ is the comoving time of the effective $\phi$--fluid ({\em 
i.e.}, the  proper time of observers comoving with this fluid and with 
four--velocity~(\ref{4velocity})) and $\Theta$ is the expansion scalar of 
the fluid \cite{Waldbook}. In a  Bianchi~I universe 
$\Theta=3H$ and $\tau=t$. Near a fixed point (which lies in the GR plane) 
${\cal KT}$ is a first--order quantity and Eq.~(\ref{KTevolution}) 
reduces, to linear order, to 
\be
\frac{ d\left( {\cal KT}\right)}{dt} =  
-3H_0 {\cal KT} -\frac{ \left( \delta \ddot{\phi} +3H_0\delta \dot{\phi} 
\right) }{8\pi \phi_0 } 
\ee  
or, in the light of the previous discussion,
\be
\frac{ d\left( {\cal KT} \right) }{dt} =  
- \frac{ 3H_0 | \delta \dot{\phi} | }{ 8\pi \phi_0}  
+ \frac{\omega_0^2 \delta \phi }{ 8\pi \phi_0 }  \,.
\ee
For expanding de Sitter spaces with purely imaginary $\omega_0$ it is 
$\omega_0^2<0$. In order for the scalar field gradient 
$\nabla^a \phi =-\dot{\phi} \, {\delta^a}_0$ to be future--oriented it 
must 
be $\dot{\phi}<0$, which implies that $\delta\phi =\delta_0 \, 
\mbox{e}^{\alpha t}<0$, or $\delta_0<0$.  Then, for the exceptional 
expanding de Sitter  fixed 
points with imaginary $\omega_0$, we have 
\begin{eqnarray}
\frac{d\left( {\cal KT} \right)}{dt} &=&  
- \frac{3H_0 \alpha |\delta \phi | }{8\pi \phi_0}  
+ \frac{\omega_0^2 \delta \phi}{8\pi \phi_0 } \nonumber\\
&&\nonumber\\
&=& \frac{ |\delta \phi|}{8\pi \phi_0} \left( -3H_0\alpha -\omega_0^2 
\right) \nonumber\\
&&\nonumber\\
& = & \frac{\alpha^2}{8\pi \phi_0} \, |\delta\phi | >0 
\end{eqnarray}
using Eq.~(\ref{algebraic}): ${\cal KT}$ always grows near these 
repellors, describing the departure from the GR equilibrium state. The 
reason for this behaviour is clearly due to the third term $\Box 
\phi/(8\pi \phi)$ in the right--hand side of Eq.~(\ref{KTevolution}).

\section{Conclusions}
\label{sec:5}
\setcounter{equation}{0}

Two basic insights have been obtained thus far in the first--order 
thermodynamics of scalar--tensor (including viable Horndeski) gravity 
\cite{Faraoni:2018qdr,Faraoni:2021lfc,Faraoni:2021jri, 
Giusti:2022tgq, Faraoni:2023hwu}. The first one is that the expansion of 
the three--space seen by observers comoving with the effective 
$\phi$--fluid ``cools'' gravity. The second is that gravity is ``hot'' 
({\em i.e.}, ${\cal KT} \to +\infty$) near spacetime singularities.

The idea that ``expansion cools gravity'' ({\em i.e.}, ${\cal KT}\to 0$) 
was deduced in 
Refs.~\cite{Faraoni:2018qdr,Faraoni:2021lfc,Faraoni:2021jri, 
Giusti:2022tgq,Faraoni:2023hwu} using situations in which $\Box\phi=0$. 
The lesson from the present study is that this statement is not always 
true when $\Box\phi \neq 0$. The term $\Box\phi/(8\pi \phi)$ in 
Eq.~(\ref{KTevolution}) cannot be expressed unambigously in terms of 
${\cal KT}$ or its powers or derivatives. This third term in the 
right--hand side of~(\ref{KTevolution}) is reminiscent of entropy 
generation terms in non--equilibrium thermodynamics and it is fair to say 
that it is the dynamics of the scalar field itself, embodied in 
$\Box\phi/\phi$, that drives gravity away from the GR equilibrium state.

When $\omega=$~const., $V(\phi) \equiv 0$, and in the presence of 
conformally invariant matter (for example, in the radiation era), it is 
$\Box\phi=0$ because Eq.~(\ref{fe2}) in the presence of matter and 
with a quadratic potential $ V(\phi)=m^2 \phi^2/2$  becomes
\be  
\square\phi = \frac{8\pi T^\mathrm{(m)} }{2\omega + 3}    
\ee
where $T^\mathrm{(m)}$ is the trace of the matter energy--momentum tensor. 
$\Box\phi$ vanishes in the presence of conformally invariant matter with 
zero trace, such as a radiation fluid in the radiation era, during which 
the expansion of space causes gravity to approach GR. This phenomenon was 
indeed reported in FLRW scalar--tensor cosmology 
\cite{Damour:1993id,Damour:1992kf}. However, was not expected in 
other cosmological eras. The convergence of scalar--tensor to GR cosmology 
has been debated at length and we hope that our approach can shed some 
light on this issue, which we will discuss in a future publication.

Another lesson garnered from the discussion of the previous section is 
that the degree of anisotropy $\Sigma/H_0^2$ commonly used in the 
literature on Bianchi universes does not tell the full story about the 
approach to, or departure from the GR state because it tends to zero for 
the exceptional expanding de Sitter fixed points with imaginary 
$\omega_0$ that are phase space repellors.

To conclude, more research is needed to understand scalar--tensor gravity 
(and even more for Horndeski gravity) from the point of view of 
first--order thermodynamics. We remind the reader that this formalism is, 
ultimately, only an analogy; nevertheless, it is proving useful from the 
theoretical point of view and it is building up to a consistent framework 
to understand at least scalar--tensor gravity in the increasingly wider 
spectrum of alternatives to GR.

\begin{acknowledgments}

V.~F. is grateful to Peter Dunsby, Andrea Giusti, Orlando Luongo, and 
Lavinia Heisenberg for discussions. This work is supported, in part, by 
the Natural Sciences \& Engineering Research Council of Canada (grant 
2023--03234 to V.~F.) and by a Bishop's University Graduate Entrance 
Scholarship (J.~H.).

\end{acknowledgments}

\begin{appendices}
\section{The pathological line $\phi=0$ in the $\dot{\phi}=0$ plane of 
the phase space}
\label{Appendix:A}
\renewcommand{\theequation}{A.\arabic{equation}}
\setcounter{equation}{0}

Let us consider an exact Bianchi~I solution of Brans--Dicke gravity that 
asymptotes to a  $\phi=0 $ solution. Assuming $V(\phi) \equiv 0$, which 
yields $\Box\phi=0$, 
consider the power--law {\it ansatz} for  the scalar field 
\be
    \phi(t) = \phi_0 t^{\alpha}
\ee
where $ \phi_0 $ is a positive constant, $t>0$, and $\alpha$ is 
assumed to be 
negative to guarantee that the gradient $\nabla^a \phi $ is 
future--oriented. The corresponding Hubble function 
\be
    H(t) = \frac{1-\alpha}{3t} 
\ee
is always positive, describing an expanding universe, and $H(t) \to 0^{+}$ 
as $t 
\to +\infty$. The shear 
\begin{gather}
    \Sigma(t) = -\frac{ \left( 3\alpha^2\omega +4\alpha ^2-2 \alpha 
-2 \right) }{12t^2}
\end{gather}
is positive if 
\begin{gather}
    \omega < -\frac{2\left( 2\alpha +1 \right)\left( \alpha -1 
\right)}{3 \alpha^2} \,.
\end{gather}
It is interesting that the quantity $\Sigma/H^2$, which measures the ratio 
of anisotropy to expansion, remains exactly constant during the evolution 
of this universe, signalling that GR (which corresponds to 
exactly vanishing $\Sigma$) is not approached. Formally, for this  
solution 
it is
\be
    {\cal KT} = \frac{|\Dot{\phi}|}{8\pi\phi} =  
 \frac{| \alpha|}{8\pi t} \to 0^{+} \quad \mbox{as  } \, t\to +\infty \,.
\ee
Although the expansion of 3--space ``cools'' this Brans--Dicke gravity, 
the zero temperature limit is not GR and is indeed a physical pathology 
corresponding to infinite $G_\mathrm{eff}$, which should be excluded from 
the range of physical possibilities. This means that a grain of salt is 
needed in the physical interpretation of the first--order thermodynamics 
of scalar--tensor gravity (which is not defined for $\phi=0$). In any 
case, the Minkowski space obtained for $V\equiv 0$, $H_0=0$, 
$\omega_0=0$ is unstable, as seen in Sec.~\ref{subsec:3A}.

\end{appendices}

\end{document}